\documentclass[aps,showpacs,prl,
reprint,
  ]{revtex4-1}
\usepackage{graphicx}
\usepackage{natbib}
\usepackage{amsmath}
\usepackage{epstopdf}
\usepackage{color}
\usepackage{upgreek}
\usepackage{gensymb}

\begin{document}


\title{Strong coupling of Rydberg atoms and surface phonon polaritons on piezoelectric superlattices}

\author{Jiteng Sheng}
\author{Yuanxi Chao}
\author{James P. Shaffer}

\affiliation{Homer L. Dodge Department of Physics and Astronomy,The University of Oklahoma, \\440 W. Brooks Street, Norman, OK 73019, USA}

\date{\today}

\begin{abstract}
We propose a hybrid quantum system where the strong coupling regime can be achieved between a Rydberg atomic ensemble and propagating surface phonon polaritons on a piezoelectric superlattice. By exploiting the large electric dipole moment and long lifetime of Rydberg atoms as well as tightly confined surface phonon polariton modes, it is possible to achieve a coupling constant far exceeding the relevant decay rates. The frequency of the surface mode can be selected so it is resonant with a Rydberg transition by engineering the piezoelectric superlattice. We describe a way to observe the Rabi splitting associated with the strong coupling regime under realistic experimental conditions. The system can be viewed as a new type of optomechanical system.
\end{abstract}


\maketitle

Atom-surface interactions continue to attract attention because they are an essential factor in many areas of physics \cite{lennard1932,wylie1984,2011casimir}. For example, recent work with surface phonon polaritons (SPhPs) has focused on realizing quantum photonic devices using atom-surface coupling \cite{shaffer2011,kubler2013}. Moreover, a great deal of effort has been invested in controlling the interaction between atoms and modified surfaces, including photonic crystals \cite{tiecke2014nanophotonic,goban2015superradiance}, nanofibers \cite{vetsch2010optical}, as well as microspheres \cite{shomroni2014all}.

Much of the work on atom-surface interaction has investigated the weak coupling regime, where the lifetime and energy of an atom can be modified by a surface \cite{barton1997,bloch2005,chance1978,kubler2010}. The strong coupling regime, where coherent interaction dominates, is more interesting because it is usually a prerequisite for quantum hybrid systems which rely on coherent control of the coupling \cite{schoelkopf2008,wallquist2009hybrid,hybridreview2015,saffmanpra2014,hafezi2012atomic,Schmiedmayer2009PRL}. However, strong atom-surface coupling is difficult to achieve due to small coupling constants and the large number of modes near the surface with which an atom can interact. Most proposals, so far, require placing atoms within a reduced wavelength of a surface ($\lambda/2 \pi$), which is technically challenging at optical wavelengths. Vacuum Rabi splitting and strong coupling have been observed for surface plasmon polaritons (SPPs) and artificial atoms, such as J-aggregates, dye molecules, and quantum dots, owing to high oscillator strength, large local field enhancement, and fixed wavevector with a directional pumping field \cite{Barnersreview,bellessa2004strong,organicsemi,6G,akimov2007generation}. Strong coupling between atoms and SPhPs has not been observed or even proposed, to our knowledge.

SPhPs are hybrid modes consisting of electromagnetic fields and crystal vibrations, typically bound to a dielectric surface. The volume of the electromagnetic field can be significantly reduced near the SPhP resonance leading to a large field enhancement. Although they attract less attention than SPPs, great progress has been made developing artificial materials that support SPhPs. For instance, low-loss materials have been fabricated for infrared SPhPs \cite{caldwell2013low,caldwell2014sub,caldwell2015review}. Microwave SPhPs can be constructed with engineered frequencies and bandwidths by introducing suitable superlattices \cite{hu2012mimicing,lu1999,chao2016}.

We propose a quantum hybrid system where strong coupling can be achieved between a Rydberg atomic ensemble and a SPhP mode on a piezoelectric superlattice (PSL) \cite{lu1999}. A PSL is a metamaterial with periodically modulated piezoelectric coefficient \cite{Ming03,hu2012mimicing}. The resonant frequencies of the SPhPs, which are usually in the microwave range, can be modified by changing the period of the PSL. Compared to SPhPs on natural materials, SPhPs on PSLs provide a more feasible platform to couple atoms to surface excitations in the near field regime. The atom-surface distances can be $~$mm instead of $\sim 100\,$nm. PSLs can be engineered, so the SPhPs are resonant with a Rydberg atom transition, which is usually impossible to do using a natural crystal.

 Rydberg atoms are highly excited atoms with huge dipole moments, $\mu$, for mm-microwave transitions. $\mu$ can be more than three orders of magnitude larger than alkali valence transitions. The large $\mu$ can partially compensate the reduction of the coupling constant due to the smaller transition frequency when compared to optical frequencies. The linewidth of the Rydberg transition can be narrow, $\sim$kHz, which results in extremely small atomic decay. The union of small atomic decay, large $\mu$ and tightly confined electromagnetical fields suggests strong coupling can be achieved with Rydberg atoms and SPhPs.

\begin{figure}[h]
\includegraphics[width=1\linewidth]{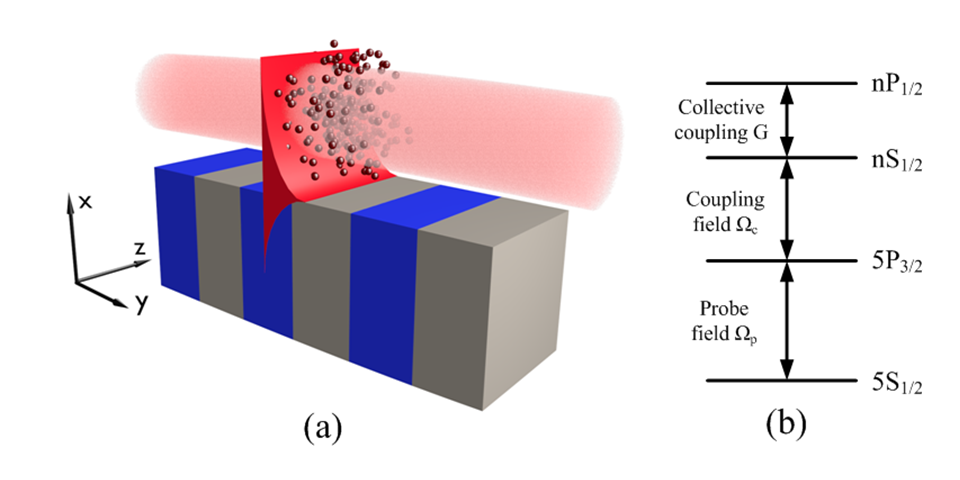}
\caption{ (a) Schematic picture of Rydberg atoms coupled to a SPhP propagating on a PPLN crystal. The laser beams parallel to the surface excite the atoms into Rydberg states. (b) The rubidium atomic energy level scheme used in this work. The SPhPs are resonant with the Rydberg state transition.}\label{fig1}
\end{figure}

Consider a sample of Rydberg atoms trapped above a semi-infinite periodically poled lithium niobate (PPLN) surface, Fig.~\ref{fig1}. PPLN is a PSL, and has been extensively used in nonlinear optics. Due to the anisotropy of PPLN, SPhPs on PPLN are more complex than SPPs. In a recent study, we demonstrated that the dielectric tensor of PPLN can be diagonalized and real SPhPs exist for the crystal orientation shown in Fig.~\ref{fig1} \cite{chao2016}. The SPhPs propagate along the y direction with surface normal x, where x, y, and z are the principal axes of lithium niobate. The electric field of the SPhP can be tightly confined at the interface between the vacuum and PPLN, and decays exponentially with distance from the surface, Fig.~\ref{fig1}.

The SPhPs can be approximated as p-polarized plane waves, i.e., the electric field lies in the x-y plane \cite{chao2016}. Thus, the vector potential of the SPhP with mode $\vec k$ can be expressed as \cite{archambault2010quantum,tame2013quantum}
\begin{equation}\label{H}
\vec{A}_{\vec k}  = \sqrt {\frac{\hbar }{{2\varepsilon _0 \omega SL}}} (\mathbf{u}_y  - \frac{{k_y }}{{k_x }}\mathbf{u}_x )a_{\vec k} e^{ik_y y - i\omega t} e^{ik_x x}  + H.c.,
\end{equation}
where $a_{\vec k}$ is the SPhP destruction operator for mode $\vec k$, S is the surface area, and L is the normalized length of the SPhP mode. $\mathbf{u}_x$ and $\mathbf{u}_y$ are unit vectors in the x and y directions, respectively. The normalized length L depends on the dielectric responses in both the propagation direction, $\varepsilon _y (\omega )$, and surface normal direction, $\varepsilon _x (\omega )$ (for details see the Supplemental Material). The effective mode volume of the SPhP field is $V = S \times L$.

The atoms are resonantly dipole coupled to the surface electric field, Fig.~\ref{fig1}. The coupling constant of a single atom and a SPhP mode $\vec k$ is
\begin{equation}\label{g}
g_{\vec k}  = \sqrt {\frac{\omega }{{2\hbar \varepsilon _0 V}}} e^{ik_x x} \vec \mu  \cdot (\mathbf{u}_y  - \frac{{k_y }}{{k_x }}\mathbf{u}_x ).
\end{equation}
The effective Hamiltonian for N atoms interacting with a quantized surface mode $\vec k$ can be written in the Tavis-Cummings form \cite{gonzalez2013theory},
\begin{equation}\label{Hal}
H = \omega _a \Sigma ^\dag  \Sigma  + \omega a_{\vec k}^\dag  a_{\vec k}  + G_{\vec k} (a_{\vec k} \Sigma ^\dag   + a_{\vec k}^\dag  \Sigma ).
\end{equation}
Here, $\omega _a$ is the atomic transition frequency, and $\Sigma$($\Sigma ^\dag$) describes the collective atomic destruction (creation) operator. The collective coupling constant for an atomic ensemble consisting of N atoms is approximately
\begin{equation}\label{G}
G_{\vec k} \approx \sqrt N g_{\vec k}.
\end{equation}
The approximation is valid when the size of the atomic sample is much smaller than the SPhP decay length along the $x$-direction, i.e. all atoms experience the same electric field. Eq.~\ref{Hal} describes the collective coupling between an atomic ensemble and a single SPhP mode.

For dipoles oriented perpendicular or parallel to the surface, as well as an isotropic average over these two orientations, factors of 1/2, $\left| {\varepsilon _y (1 - \varepsilon _x )/(1 - \varepsilon _y )} \right|$, and $(1 + \left| {\varepsilon _y (1 - \varepsilon _x )/(1 - \varepsilon _y )} \right|)/3$ have to be included to calculate the normalized length, L. $\varepsilon _i$, $i=x,y,z$, depends on $\omega$ but we have suppressed the dependence to make the formulas clear. The associated g-factors are
\begin{equation}
g_{||}  = \mu \sqrt {\frac{\omega }{{4\hbar \varepsilon _0 V}}} e^{ik_x x},
\end{equation}
\begin{equation}
g_ \bot   = \mu \sqrt {\frac{\omega }{{2\hbar \varepsilon _0 V}}\left| {\frac{{\varepsilon _y (1 - \varepsilon _x )}}{{(1 - \varepsilon _y )}}} \right|} e^{ik_x x},
\end{equation}
and
\begin{equation}
g_{iso}  = \mu \sqrt {\frac{\omega }{{6\hbar \varepsilon _0 V}}\left( {1 + \left| {\frac{{\varepsilon _y (1 - \varepsilon _x )}}{{(1 - \varepsilon _y )}}} \right|} \right)} e^{ik_x x}.
\end{equation}
The orientations of the Rydberg atoms can be controlled by external fields, e.g. electric fields.

\begin{figure}[h]
\includegraphics[width=1\linewidth]{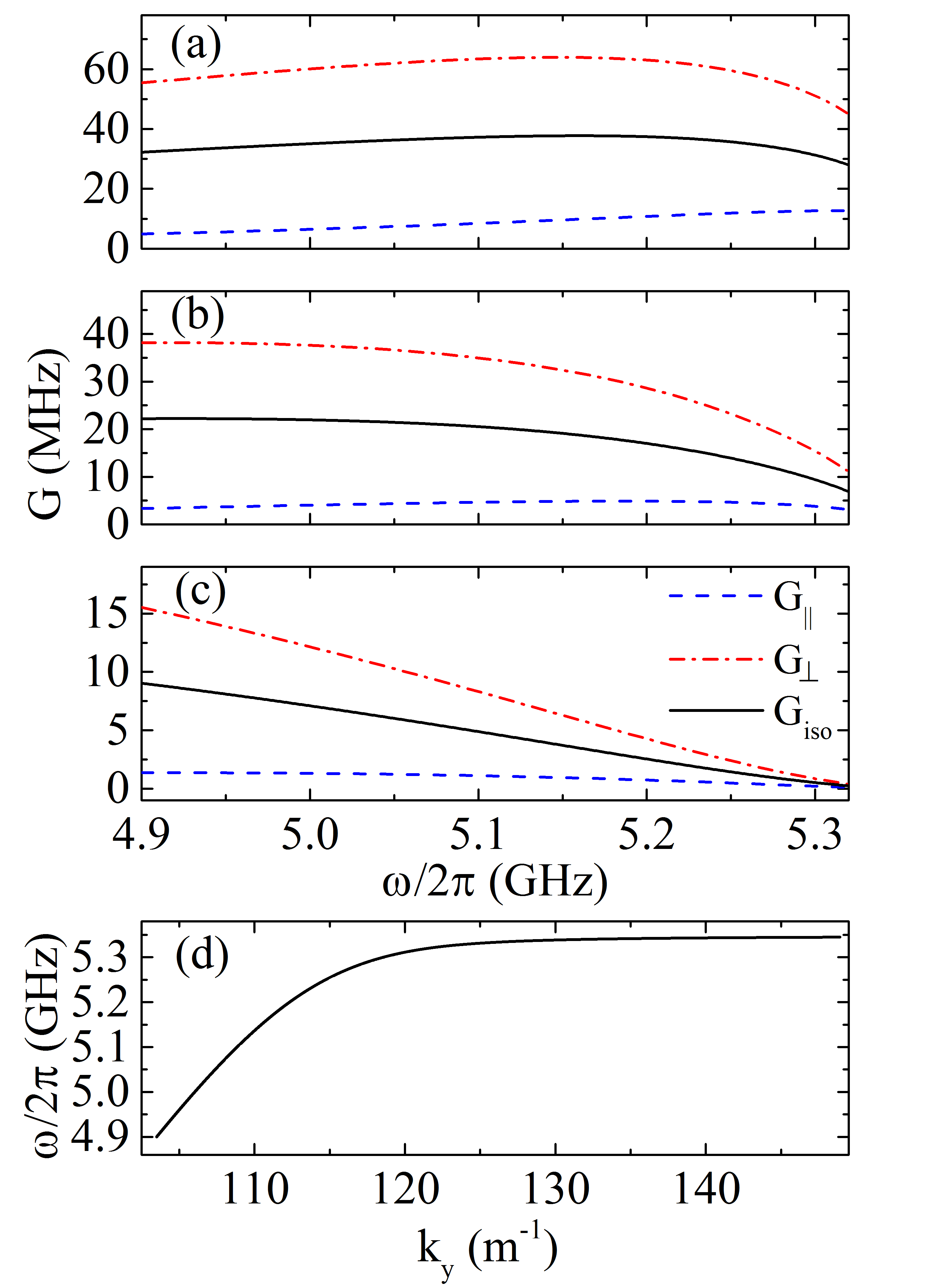}
\caption{Coupling constants for parallel ($\|$), perpendicular ($\bot$), and isotropic ($iso$) orientations of Rydberg atoms at different distances (a) 1 mm, (b) 30 mm, and (c) 100 mm, from the PSL surface, respectively. (d) shows the dispersion curve for the SPhP with parameters as described in the text. k$_\mathrm{y}$ is the propagation constant along the y-direction of the crystal, shown in Fig.~\ref{fig1}.}\label{fig2}
\end{figure}

If we choose the period of the PPLN superlattice to be $\sim$ 1 $\mu$m, then there is a bandgap between 4.9 and 5.3 GHz, which is the frequency range for a real SPhP with resonance frequency near 5.3 GHz \cite{chao2016}. PPLN superlattices with such small periods are possible to construct with modern fabrication technologies, e.g., the direct-write e-beam method \cite{son2005direct}. The frequency of the SPhP matches the 87S$_{1/2}$ to 87P$_{1/2}$ Rydberg transition of rubidium with $\mu \sim$ 8000ea$_0$. The $G$ for each orientation is plotted in Fig.~\ref{fig2}(a-c) at different distances away from the surface. As shown in Fig.~\ref{fig2}(a), the coupling constant can be larger than 50$\,$MHz for a Rydberg atomic ensemble trapped 1$\,$mm away from the surface in a volume of $2 \times 2 \times 1\,$mm$^3$ with a Rydberg atom density $n = 1 \times 10^9\,$cm$^{-3}$. The PPLN surface is 1$\,$mm wide and 5$\,$cm long for this calculation. In order to reach the strong coupling regime, a dilute atomic ensemble with an average distance between atoms larger than 10 $\mu$m is considered, so that the Rydberg atom interactions \cite{AAMOP2014} and Rydberg atom molecules \cite{Bendkowsky2009,Overstreet2009,Schwettmann2007,Greene2001,Cote2002} can be ignored at the level of the present calculation. Rydberg blockade \cite{saffman2010quantum,comparat2010dipole} does not destroy the effect, but will limit the number of Rydberg excitations present in the sample. Future investigations will use the blockade effect to enable single collective atomic excitations to interact with SPhPs, but these considerations are beyond the scope of the current paper. $G$ can be increased by fabricating the metamaterial surface as a waveguide or resonator \cite{hummer2013weak}.

To achieve strong coupling, $G$ must be larger than the dissipation present in the system \cite{Barnersreview,gonzalez2013theory}, which is mainly from the atomic, radiative decay, $\gamma_a$, and the decay of the SPhP mode propagating on the surface due to the loss into the crystal bulk polariton modes, $\gamma_{spp}$. The Rydberg atoms have a lifetime of $\sim 1 \,$ms, corresponding to a decay $\gamma_a \sim 1 \,$kHz. The decay of the SPhP mode is frequency dependent. The SPhP decay has a maximum value when the frequency is at the SPhP resonance. At this point, the SPhP decay is equal to the damping constant of the crystal, $\Gamma$, and decreases as the frequency is detuned to the red side of the resonance \cite{nkoma1974}. The change in the decay constant is typically less than 1 order of magnitude. The resonant damping constant is $\Gamma \sim 0.001 \omega_0$ for lithium niobate, where $\omega_0$ is the SPhP resonance frequency \cite{yin2011polaritons}. Using this estimate, the decay of the SPhP is $\gamma_{spp} \sim 5\,$MHz at its peak. Hence, the strong coupling condition is satisfied as G$>> \gamma_a$, $\gamma_{spp}$, and the Rabi splitting at resonance is $\sqrt {G^2  - (\gamma _a  - \gamma _{spp} )^2 /4}\approx G$.

The SPhPs have a broad bandwidth compared to atomic decay. As shown in Fig.~\ref{fig2}(a-c), $G$ is relatively uniform over $\sim 400\,$MHz. Consequently, the SPhP can be resonant with the Rydberg transition, as typical cavity quantum electrodynamics systems require, over a relatively large energy range. The broad nature of the resonance is in sharp contrast with the narrow bandwidth of Rydberg atom coupling to on chip, microwave resonators. The large, uniform coupling bandwidth is advantageous for making the PSL because more error can be tolerated in the period of the superlattice. Having a relatively narrow atomic transition to couple to the SPhP also has advantages. Fig.~\ref{fig2}(d) shows the dispersion curve for the SPhP described in this paper. There are many modes with different propagation constants shown in Fig.~\ref{fig2}(d), but energy conservation associated with the narrow Rydberg transition picks out a small range of k$_\mathrm{y}$, except at resonance, and the collective atom-SPhP coupling can be effectively described by the Hamiltonian in Eq.~\ref{Hal}.

It is possible to drive a particular SPhP mode with a microwave field. In this case, the SPhPs that are excited can produce a continuous wave electric field near the surface that can interact with the atoms. A SPhP mode with fixed wavevector can be excited, for example, by applying the edge-coupling method \cite{huber2008focusing}. When the frequency of the pumping field is tuned near the atomic resonance frequency, one can observe phenomenon similar to those occurring when the cavity frequency is tuned in coupled atom-cavity systems \cite{bellessa2004strong}.

\begin{figure}[h]
\includegraphics[width=1\linewidth]{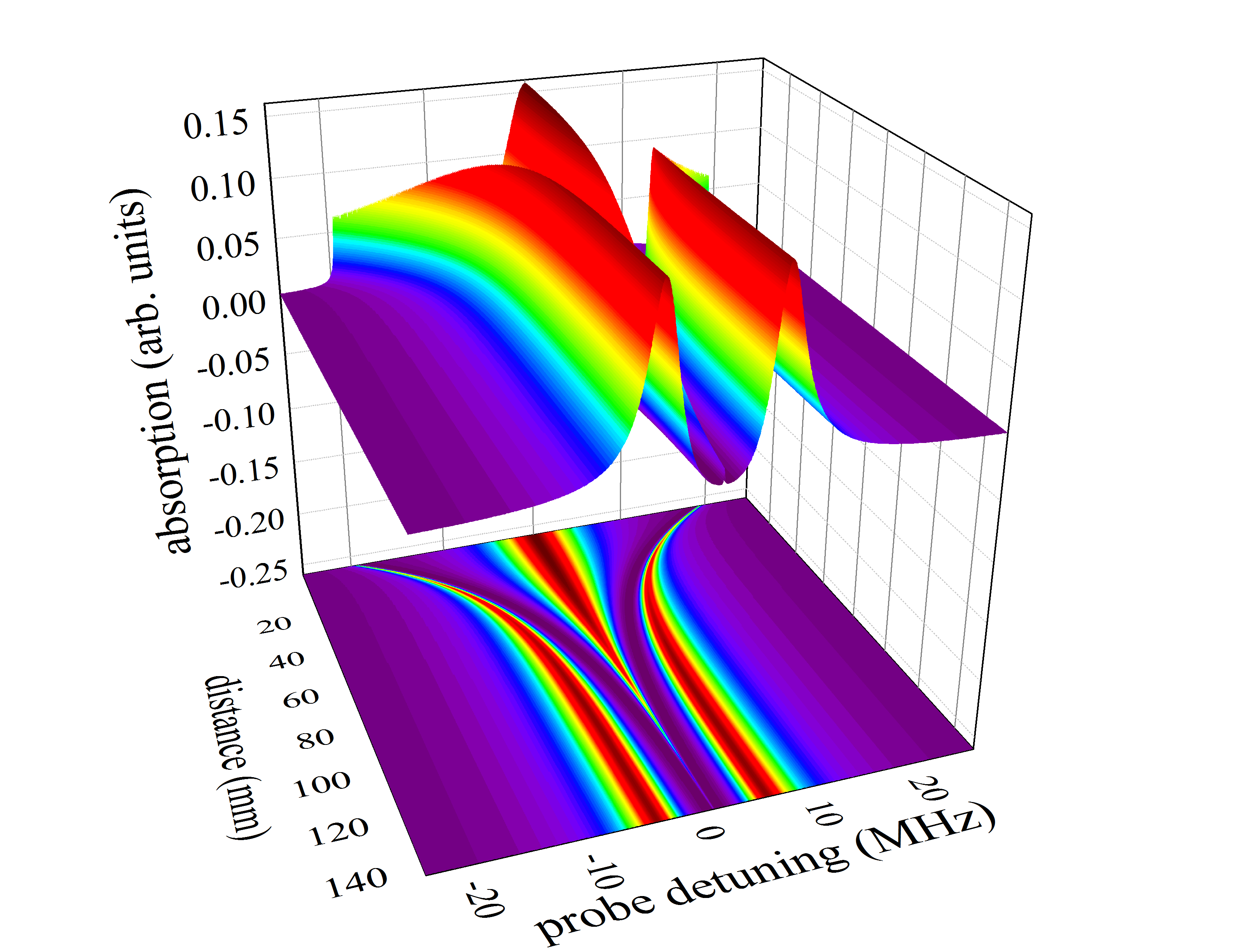}
\caption{Absorption spectrum for the atoms at different distances away from the surface. The probe and coupling laser Rabi frequencies are $1\,$MHz and $10\,$MHz, respectively. The PSL and SPhP are the same as for the other calculations.}\label{fig3}
\end{figure}

 One approach to observe the Rabi splitting resulting from the strong coupling between Rydberg atoms and SPhPs, inspired by the demonstrations of strong coupling in SPP systems \cite{bellessa2004strong}, is to use the SPhPs driven by a microwave field, as described in the prior paragraph. Similar to experiments for measuring microwave power with Rydberg atoms \cite{sedlacek2012microwave,fan2015atom}, Fig.~\ref{fig1}, electromagnetically induced transparency (EIT) \cite{fleischhauer2005,acell} can be used to measure the strong coupling between the atoms and the SPhP mode. By replacing the microwave field with the collective coupling between the atoms and the SPhP mode, the collective atom-SPhP coupling constant can be measured. A similar idea has been predicted \cite{field1993vacuum,rice1996cavity} and observed \cite{tanji2011vacuum} in a lambda-type EIT cavity system.

 To observe the collective Rydberg atom-SPhP coupling, rubidium atoms in a magneto-optical trap (MOT) can be placed a few millimeters away from a PPLN surface. A probe laser drives the 5S$_{1/2}$ to 5P$_{3/2}$ transition and a coupling laser drives the 5P$_{3/2}$ to 87S$_{1/2}$ transition. The probe and coupling laser beams overlap the MOT. A similar experimental setup has been used to detect the electric field near the surface of single crystal quartz at atom surface separations $< 50\,\mu$m \cite{sedlacek2015electric}. The transmission spectrum of the probe laser then depends on the separation between the atoms and the surface which can realistically be varied on $10\,\mu$m scales. Fig.~\ref{fig3} shows the probe laser absorption spectra for atoms at various distances away from the PSL surface. The transparency window, between the two absorption peaks, splits into two when the atoms are moved close to the surface. The Rabi frequency of the probe (coupling) laser used in Fig.~\ref{fig3} is chosen to be $1\,$MHz (10$\,$MHz), and $G$ for the neighboring Rydberg transition varies from $\sim 1 - 40\,$MHz when the atoms are moved from 150$\,$mm to 1$\,$mm away from the surface. The motion of the atoms is neglected since the temperature of the atoms is $\sim 100\,\mu K$ and the decay length of the SPhP is large compared to the size of the atomic sample.

 It is also possible to detect the atom-SPhP coupling by observing the collective decay of the atomic sample as influenced by the presence of the SPhP modes \cite{goban2015superradiance,chang2006quantum}, i.e. the enhanced atomic decay will spectrally broaden the EIT lineshape. By analyzing the Purcell factor, which is defined by the ratio of the atomic decay into the SPhP modes to decay into other modes, we find the enhancement of the atomic decay can be much larger than 1 when the frequency is close to, but not at, the SPhP resonance. $G >> \gamma_a$, $\gamma_{spp}$ and the SPhP is unbound in the $y$-direction since it can propagate away from the atoms.

\begin{figure}[h]
\includegraphics[width=0.9\linewidth]{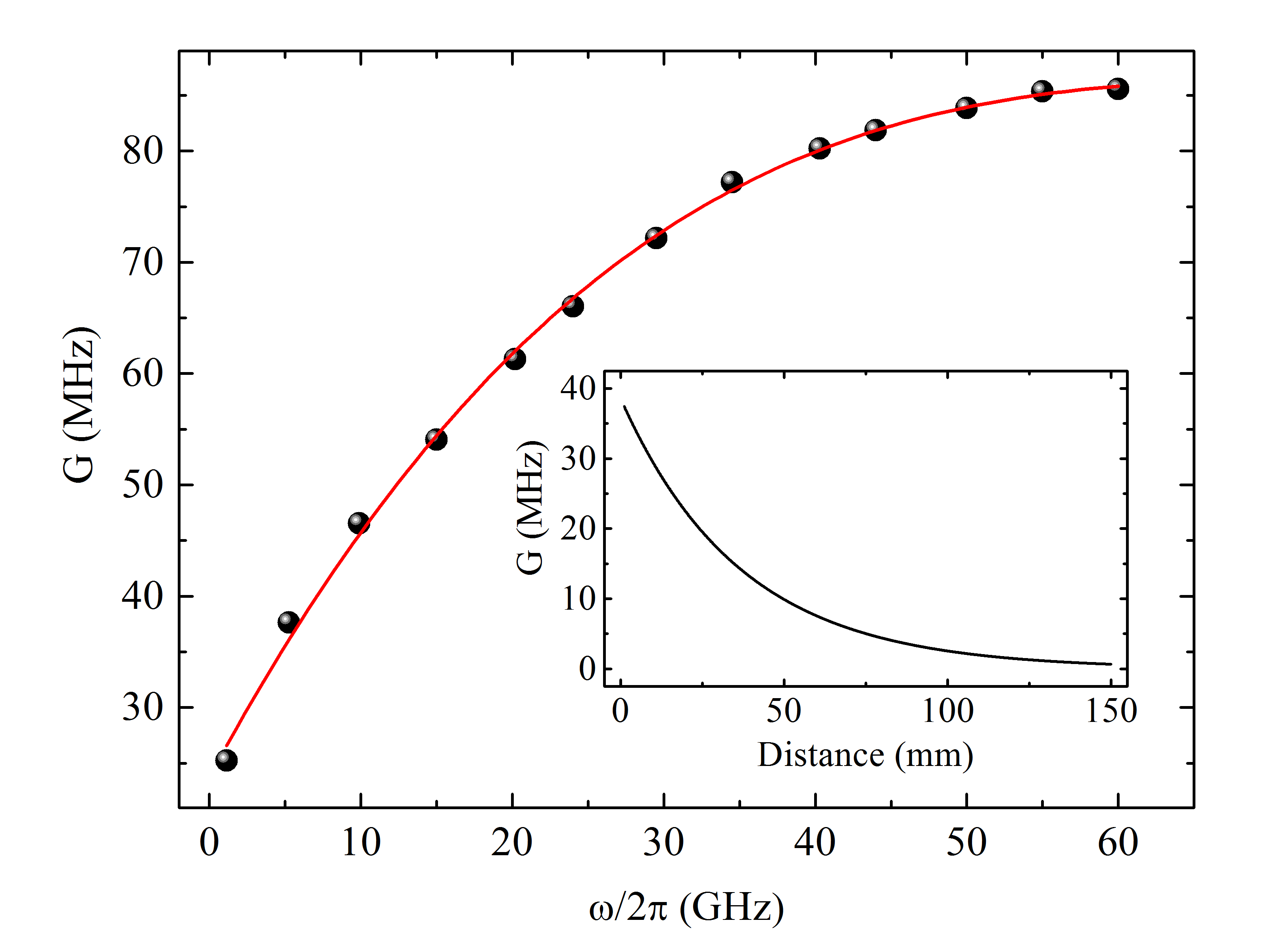}
\caption{$G$ as a function of SPhP frequency. The inset shows $G$ as a function of the distance away from the surface at a SPhP frequency around 5 GHz. The red line is a guide line.}\label{fig4}
\end{figure}

The SPhP frequencies can be modified by engineering the period of the PSL. Fig.~\ref{fig4} shows $G$ as a function of the SPhP frequency. In Fig.~\ref{fig4}, the atoms are assumed to be in the near field of the surface, i.e. the exponential term in Eq.~\ref{g} is ignored. Each data point in Fig.~\ref{fig4} corresponds to a PSL with a different period and a Rydberg atom transition at a different principal quantum number, $n$. In Fig.~\ref{fig4}, the coupling constant increases as the frequency increases. The increase can be attributed to $G$ increasing as $\omega$ and L decreasing as $\omega$. In order to be resonant with higher frequency SPhPs, a lower principle number $n$ of the Rydberg atom is used. $\mu$ is proportional to $n^2$ so as $\omega$ increases $\mu$ decreases as $n^2$.

Most previous studies of atom-surface interaction use low-lying energy levels, with optical or infrared transitions \cite{stehle2011plasmonically,laliotis2014casimir} making it necessary to position the atoms within $100\,$nm of a surface. Microwave strip lines and Rydberg atoms require $\sim 50\,\mu$m atom-surface separations. SPhPs on PSLs can potentially operate from GHz-THz and have near field coupling ranges of cm-$\mu$m. The electric field gradients for SPhPs are not as large as for superconducting, on chip cavities. SPhPs also do not couple to free space radiation modes, since those modes cannot simultaneously meet both energy and momentum conservation. Unlike most natural materials, PSL's can also support longitudinal SPhPs \cite{Zhu2004}. Structures that can couple surface modes to bulk propagating modes can be designed \cite{Zhou2012}. The fact that the atoms can be placed relatively far from the surface can enable one to interface them with superconducting qubits without having the light destroy the superconductivity. These properties can be important for frequency conversion at the single quantum level, interfaces to quantum based devices, and optomechanical transduction. Coherent control of atom-SPhP coupling can be an important part of the quantum engineering toolbox.


In conclusion, we have suggested a collective atom-surface quantum hybrid system involving a Rydberg atomic ensemble coupled to a propagating SPhP mode on a PSL. We demonstrated that strong coupling can be achieved by properly engineering a PSL. The system provides an interface that can allow the transport of quantum information between a high and a low temperature environment and can serve as an atomic interface to polaritronic and microwave devices, including those operating at the quantum level. The system also presents the possibility to study the optomechanics of the atom-surface system, i.e. the atoms interact with the crystal vibrations through the electro-magnetic field. Finally, the proposed system is more experimentally forgiving than atom-SPP ones and can be used for proof of principle experiments for atom-SPP applications.

\section{Acknowledgments}
This work was supported by the AFOSR (FA9550-12-1-0282) and NSF (PHY-1104424).

\bibliographystyle{apsrev4-1}
%

\end{document}